\documentclass[aps,10pt,prd,twocolumn,preprintnumbers,amsmath,amssymb,nofootinbib,superscriptaddress,a4paper,showpacs]{revtex4-1}
\usepackage[breaklinks]{hyperref}
\usepackage[latin1]{inputenc}
\usepackage{graphicx}
\usepackage{color}
\usepackage{amsfonts,amsthm}
\usepackage{bm,bbm}
\usepackage[OT2,OT1]{fontenc}
\newcommand\cyr{%
\renewcommand\rmdefault{wncyr}%
\renewcommand\sfdefault{wncyss}%
\renewcommand\encodingdefault{OT2}%
\normalfont
\selectfont}
\DeclareTextFontCommand{\textcyr}{\cyr}

\def\be{\begin{equation}}
\def\ee{\end{equation}}
\def\ba{\begin{eqnarray}}
\def\ea{\end{eqnarray}}
\def\bs{\begin{subequations}}
\def\es{\end{subequations}}
\def\a{\alpha}

\def\de{\delta}
\def\g{\gamma}
\def\la{\lambda}
\def\k{\kappa}

\def\om{\omega}

\def\s{\sigma}

\def\vp{\varphi}

\def\bu{\bar{\mu}}
\def\cA{\mathcal{A}}

\def\cC{\mathcal{C}}

\def\cE{\mathcal{E}}

\def\cH{\mathcal{H}}
\def\cI{\mathcal{I}}
\def\cJ{\mathcal{J}}
\def\cK{\mathcal{K}}

\def\cN{\mathcal{N}}
\def\cO{\mathcal{O}}

\def\cV{\mathcal{V}}

\def\p{\partial}

\newcommand{\Eq}[1]{(\ref{#1})}
\def\com{\color{magenta}}
\def\cob{\color{blue}}

\newcommand{\arX}[1]{\href{http://arxiv.org/abs/#1}{{\ttfamily\com arXiv:#1}}}
\newcommand{\doin}[6]{\href{http://dx.doi.org/#1}{{\cob #2 #3 {\bf #4}, #5 (#6)}}}
\newcommand{\doinn}[5]{\href{http://dx.doi.org/#1}{{\cob #2 {\bf #3}, #4 (#5)}}}
\newcommand{\doij}[5]{\href{http://dx.doi.org/#1}{{\cob #2 #3 (#5) #4}}}

\newcommand{\procs}[6]{in \emph{#1}, edited by #2 (#3, #4, #5, #6)} 
\newcommand{\procsin}[5]{in \emph{#1}, edited by #2 (#3, #4, #5)} % for Singapore
\newcommand{\tia}[1]{}

\def\lp{\ell_{\rm Pl}}

\def\rme{e}
\def\rmd{d}
\def\rmi{i}

\begin{document}

\title{Loop quantum cosmology from group field theory}

\author{Gianluca Calcagni}
%\email{calcagni@iem.cfmac.csic.es}
\affiliation{Instituto de Estructura de la Materia, CSIC, Serrano 121, 28006 Madrid, Spain}

\date{\small July 30, 2014}

\begin{abstract}
We show that the effective dynamics of the recently proposed isotropic condensate state of group field theory with Laplacian kinetic operator can be equivalent to that of homogeneous and isotropic loop quantum cosmology in the improved dynamics quantization scheme, where the area of elementary holonomy plaquettes is constant. This constitutes a somewhat surprising example of a cosmological model of quantum gravity where the operations of minisuperspace symmetry reduction and quantization can actually commute.
\end{abstract}


% 04.60.Pp, Loop quantum gravity, quantum geometry, spin  foams
% 04.60.-m Quantum gravity
% 98.80.Qc	Quantum cosmology 

\preprint{\doin{10.1103/PhysRevD.90.064047}{Phys.\ Rev.}{D}{90}{064047}{2014} \hspace{10.5cm} \arX{1407.8166}}

\maketitle

%%%%%%%%%%%%%%%%%%%%%%%%%%%%%%%%%%%%%%%%%%%%%%%%%%%%%%%%%%%%%%%%%%%%%%%%%%%%%%%%%%%%%%%%%
%%%%%%%%%%%%%%%%%%%%%%%%%%%%%%%%%%%%%%%%%%%%%%%%%%%%%%%%%%%%%%%%%%%%%%%%%%%%%%%%%%%%%%%%%

\section{Introduction} 

Hamiltonian theories of quantum gravity are notoriously hard to tame. Apparently inexhaustible technical difficulties make the solution of background-independent canonical quantum constraints a challenging task. However, if one restricts the theory to homogeneous geometries, one can construct a quantum model on a finite-dimensional minisuperspace \cite{De671}. It is generally agreed that minisuperspace quantization is only a toy model of a quantum theory of geometry. The symmetry reduction is performed before solving the quantum constraints: one first restricts the classical theory to a homogeneous background and then quantizes. Dynamics is then expressed by a quantum super-Hamiltonian constraint (Wheeler--DeWitt equation) $\hat\cH\Psi=0$, where the wave function $\Psi$ depends only on geometry variables (the scale factor $a$ of the universe) and matter fields. On the other hand, a complete theory should implement quantization in a background-independent way and \emph{then} specialize, if desired, to cosmology. Since symmetry reduction and quantization do not commute in general \cite{KuR1}, %the best one can usually hope for is that
 minisuperspace models can capture only some of the qualitative features of the cosmology of the full theory.

For example, homogeneous loop quantum cosmology (LQC) is not the cosmological limit of full loop quantum gravity (LQG) but a minisuperspace model employing LQG techniques (but see the recent results of \cite{AlCi4}). Taking the expectation value of the symmetry-reduced quantum operator $\hat\cH$ on a semiclassical state peaked at classical values of the canonical variables, for a massless scalar field one gets the effective Friedmann equation (e.g., \cite{lqcr,CQC})
\be\label{Hg2}
\hspace{-0.5cm} \a\sin^2 (\bu c)= \frac{\rho}{\rho_*},\quad \rho=\frac{\dot\phi^2}{2\nu},\quad \rho_* := \frac{3}{\cV_0^{2/3}\g^2\k^2\bu^2 a^2},
\ee
where $\a=1+O(a^{-\s})$ and $\nu=1+O(a^{-\s})$ contain small corrections to inverse-volume operators, 
\be\label{bu2}
\bu =\bu(a)\propto a^{-2n}\,,\qquad n>0\,,
\ee
is an arbitrary function of the scale factor (but $n$ must be positive for internal consistency of the model), $c=c(t)$ is the Ashtekar--Barbero connection in homogeneous isotropic spacetime, $A_\a^i\propto c\,\, \de_\a^i$, Roman indices $i=1,2,3$ run over the internal gauge space, $\a=1,2,3$ are spatial indices, and $\cV_0$ is the comoving fiducial volume on which the Hamiltonian is defined. Classically, $c=\g\dot a+\textsc{k}$ ($\textsc{k}=0,\pm 1$ is the intrinsic curvature). The critical density $\rho_*$, which depends on the gravitational constant $\k^2=8\pi G$ and on the Barbero--Immirzi parameter $\g=O(1)$, is constant only in the so-called improved quantization scheme \cite{APS}
\be\label{gftqs}
\bu\propto a^{-1}\,,\qquad n=\frac12\,.%\frac{1}{a}\,.
\ee
In this case, the classical big-bang singularity is replaced (in the homogeneous theory) by a quantum bounce slightly below the Planck density. Several quantization ambiguities are present in the functions $\a$, $\nu$, and $\bu$ which cannot be fixed in the theory purely from the Friedmann equation. This is a problem typical of canonical quantization, where $\hat \cH$ can be defined by an infinite number of operator orderings. In the general case of lattice refinement \cite{bo609}, one identifies $\bu$ with the number $\mu=\ell_0\cV_0^{-1/3}=\cN^{-1/3}$ of elementary holonomy cells of comoving length size $\ell_0$ (parallel transport of the connection along closed paths) per fiducial volume. Equation \Eq{gftqs} then implies that the cell area $(a\ell_0)^2$ is constant and proportional to the Planck area $\lp^2$, but since we are ignorant about the dynamical evolution of $\cN$ in a state of the full theory, the exponent $n$ in $\bu$ can take any positive value, and, however well motivate the improved scheme \Eq{gftqs} is \cite{CoSi3}, it remains only one among many possibilities.

In this paper, we partially solve this drawback of LQC by deriving the structure of its homogeneous and isotropic dynamics \Eq{Hg2} from a fundamental quantum theory of gravity, group field theory (GFT). GFT \cite{gft} is a background-independent nonperturbative quantum setting where a special physical state can be chosen so that the conditions of homogeneity and isotropy emerge in the semiclassical continuum limit. One can then extract cosmology directly from the full quantum theory \cite{GOS1,Gie13}. The present stage of development of the model does not warrant a thorough comparison with the findings of LQC, and we are unable to check \emph{all} of its known features (such as the form of inverse-volume corrections and whether these are artifacts of the symmetry reduction). However, we remove the quantization ambiguity on $\bu$ unavoidable in the Hamiltonian formalism and argue in favor of the improved quantization scheme.

%%%%%%%%%%%%%%%%%%%%%%%%%%%%%%%%%%%%%%%%%%%%%%%%%%%%%%%%%%%%%%%%%%%%%%%%%%%%%%%%%%%%%%%%%
%%%%%%%%%%%%%%%%%%%%%%%%%%%%%%%%%%%%%%%%%%%%%%%%%%%%%%%%%%%%%%%%%%%%%%%%%%%%%%%%%%%%%%%%%

\section{GFT condensate} 

We first review group field theory and its cosmology in four dimensions \cite{GOS1,Gie13}. GFTs are quantum field theories on group manifolds. One has a complex-valued object $\vp(g):=\vp(g_1,\dots,g_4)$ dependent on four elements $g_I$ of the local gauge group $G$ of gravity. Gauge invariance of vertices is expressed by the property $\vp(g_I)=\vp(g_Ih)$ for all $h\in G$ and all $I$. The classical dynamics is governed by the action
\be\label{gftS}
S_\textsc{gft}=\int_G\rmd^4g\left[\int_G\rmd^4g'\,\vp^*(g)\cK(g,g')\,\vp(g')+V\right],
\ee
where the kinetic operator $\cK$ is an operator on $G^2$ and the potential $V=V[\vp(g),\vp^*(g)]$ is a nonlinear interaction of the fields; choices of $\cK$ and $V$ fix the model.

The classical field $\vp(g)$ is interpreted as the four-valent vertex of a spin network, with the group label $g_I$ being the holonomy of the connection along the $I$th link. To each vertex in a spin network there corresponds a $3$-simplex (a tetrahedron) in the dual simplicial complex. In this representation, $\vp(g)$ is a $3$-simplex whose four $2$-faces are labeled by the $g_I$s and $\vp(g_I)=\vp(g_Ih)$ is called closure constraint, since it is equivalent to the requirement that the four triangles close to form the tetrahedron. The interaction term $V$ describes how tetrahedra are glued together along their faces to form a $4$-simplex.

When GFTs are constructed as the generalization of loop quantum gravity, the group is $G=SU(2)$, and the geometry described by the states is three dimensional and spatial. The quantum scalar field $\hat\vp$ is expanded in terms of creation and annihilation operators on a Fock space. The field obeys the algebra $[\hat\vp(g),\,\hat\vp^\dagger(g')]=\mathbbm{1}_G(g,g')$, $[\hat\vp,\,\hat\vp]=0=[\hat\vp^\dagger,\,\hat\vp^\dagger]$, where $\mathbbm{1}_G$ is an identity operator compatible with the closure constraint. The Fock vacuum $|\emptyset\rangle$ is by definition annihilated by $\hat\vp$, $\hat\vp|\emptyset\rangle=0$, and corresponds to a ``no-spacetime'' configuration where no quantum-geometry degree of freedom is present and all area and volume operators have vanishing expectation value. By convention, $\langle\emptyset|\emptyset\rangle=1$. The one-particle GFT Fock state $|g\rangle:=\hat\vp^\dagger(g)|\emptyset\rangle$ is interpreted as the creation of a four-valent spin-network vertex or of its dual tetrahedron with labels $g_I$.

Matter fields can be added to the picture, either as emergent degrees of freedom or by hand as new coordinates in an extension of the group manifold $G^4$ \cite{ORy}. In the second case and for a real scalar field $\phi$, the GFT field becomes $\vp(g)\to\vp(g,\phi)$. The generalization of the action \Eq{gftS} and of the rest of the theory is straightforward. %We will stick with the second possibility. Adding a scalar field by hand in GFT is expected to be the exact counterpart of adding a scalar field by hand in LQC (and a cosmological GFT model with emergent matter will most likely deviate from the dynamics of LQC).

In GFT, the double limit of the continuum and semiclassicality is delicate because the pregeometric discrete structure contains different information with respect to gravity on a fixed topology. To translate statements regarding differential manifolds (homogeneity, for instance) into the language of simplicial complexes, it should be made possible to embed any such complex into a smooth continuous geometry. In a limit, if existing, where the complex describes a differentiable spatial hypersurface, each tetrahedron should be nearly flat compared to the overall curvature radius of the embedding geometry. This flatness condition is an important self-consistency check to be done at the end.

With this caveat on board, one can construct GFT quantum states in the Fock space capable of describing geometries with a continuum homogeneous limit. In a homogeneous manifold, all points of space carry the same information on the metric or the connection. In a classical dual complex where the flatness condition holds, the equivalent of points of space are tetrahedra and their metric information is carried by their group or algebra labels. The redundancy required by homogeneity is thus achieved by asking (i) that all of the building blocks of the combinatorial structure be in the same microscopic configuration and (ii) that this configuration admit a ``macroscopic'' description as a whole. This does not mean that the labels of each classical building block are fixed to the same values, which would correspond to an \emph{ad hoc} minisuperspace approximation before quantizing. Rather, one considers the $\cN$-particle state $\hat\xi^\cN|\emptyset\rangle = \hat\xi\cdots\hat\xi|\emptyset\rangle$ built from some operator $\hat\xi$ composed of creation operators summed over all possible group configurations: $\hat\xi:=\int\rmd^4g\,\xi(g)\,\hat\vp^\dagger(g)$, where the weight $\xi$ is called $\sigma$ in \cite{GOS1,Gie13}. Here we take the case of elementary building blocks, i.e., tetrahedra, but one can obtain other types of states made of ``molecular'' composites.

The quantum geometry found at these $\cN$ distinct tetrahedra is the same. Taking the limit $\cN\to+\infty$, the complex approximates to a continuum. This limit is not just formal and can be realized by a concrete physical state. In fact, a configuration with (i) an infinite number of particles in the same microscopic quantum state and (ii) characterized by one macroscopic description is nothing but a condensate. According to the lore of condensed matter, the ensemble of tetrahedra is thus represented by the gauge-invariant kinematical state
\be\label{xist}
|\xi\rangle := A\,\rme^{\hat\xi}|\emptyset\rangle\,,
\ee
where $A$ is some normalization here chosen so that $\langle\xi|\xi\rangle=1$. An easy calculation shows that $|\xi\rangle$ is a coherent state, that is, $\hat\vp(g)|\xi\rangle=\xi(g)|\xi\rangle$. The metric can be reconstructed from $\xi$ when working in momentum space.

Expression \Eq{xist} defines a nonperturbative vacuum on the kinematical Fock space, on which the GFT field acquires the nontrivial expectation value $\langle\xi|\hat\vp|\xi\rangle=\xi\neq 0$. We enforce a mean-field approximation on the condensate \Eq{xist}, expanding the field $\vp=\xi+\de\vp$ around its nonzero vacuum expectation value and truncating the equations of motion up to some order in the fluctuations $\de\vp$. Full quantum dynamics is given by the infinite tower of constraints $\langle\xi|\hat\cO\hat\cC|\xi\rangle=0$, where $\hat\cC:=\int\rmd^4g'\,\cK(g,g')\hat\vp(g')+\de \hat V/\de\hat\vp^\dagger(g)$ is the quantum version of the classical equation of motion and $\hat\cO[\hat\vp,\hat\vp^\dagger]$ is an arbitrary operator of the GFT field. Exact solutions to the quantum dynamics solve all of these conditions simultaneously. Approximated solutions can be found by imposing only the first of such constraints ($\cO=\mathbbm{1}$), which is the analogue of the Gross--Pitaevskii equation for Bose--Einstein condensation \cite{PiS}. Taking a normal ordering in $\hat V$ such that all $\hat\vp^\dagger$ are to the left of all of the $\hat\vp$, one has
\be\label{gfteomqs}
0=\langle\xi|\hat\cC|\xi\rangle=\int\rmd^4g'\,\cK(g,g')\xi(g')+\left.\frac{\de V}{\de\vp^*(g)}\right|_{\vp=\xi}\,.
\ee
Solutions $\xi(g)$ of this equation give, when plugged into Eq.\ \Eq{xist}, approximate physical states.

The scalar weight $\xi$ is interpreted as a probability distribution on the space of homogeneous geometries. It is not a wave function of the quantum geometry in the canonical sense since \Eq{gfteomqs} is nonlinear in general. In Wheeler--DeWitt and loop quantum cosmology, a wave function $\Psi$ describes a single quantum universe with fixed topology. In GFT cosmology (or GFC in short), the scalar $\xi$ is a highly quantum object, the interpretation of a continuum geometry and the semiclassical limit being recovered only by the macroscopic, large-scale collective behavior of this many-particle ensemble. In the case of molecular condensates, $\xi$ carries also information on the correlation between different quanta.

To study a concrete model of quantum dynamics, one must make a choice of operators in Eq.\ \Eq{gfteomqs}. Renormalization analyses indicate that finiteness of the theory requires the kinetic operator $\cK$ to include the Laplacian $\Delta_{g}$ on the group manifold \cite{GeBo}. Assuming that nonlinear interactions are negligible ($V=M^2|\vp|^2$) and including a matter scalar field, the dynamical equation to solve is
\be\label{gfteomqs5}
\left(\sum_{I=1}^4\Delta_{g_I}+12\cE^2\p^2_\phi+M^2\right)\xi(g,\phi)=0\,,
\ee
where $M$ is dimensionless and $\cE^2$ is some constant whose sign will be chosen later in relation with the classical equations of motion. At first we ignore matter, $\cE^2=0$. We use the irreducible representation of $SU(2)$ in a neighborhood of the identity in terms of the Pauli generators $\tau_i=\s_i/(2\rmi)$ of the $su(2)$ Lie algebra where $g_I(\pi_I)=\sqrt{1-\vec\pi_I\cdot\vec\pi_I}\mathbbm{1}_2+2\vec\tau\cdot\vec\pi_I$ is a $2\times 2$ matrix and the four three-vectors $\vec\pi_I$ are elements of $su(2)$ such that $|\vec{\pi}_I|\leqslant 1$. Thanks to gauge invariance and the closure condition, one can manipulate the dynamical equation to express it only in terms of the elements $\vec\pi_\cI$ of the first three links (dual faces). The $\vec\pi_\cI$s can be combined into the matrix invariants $\pi_{\cI\cJ}:=\vec\pi(g_\cI g_4^{-1})\cdot \vec\pi(g_\cJ g _4^{-1})$, where $\cI,\cJ=1,2,3$, $|\pi_{\cI\cJ}|\leqslant 1$ and $\pi_{\cI\cI}\geqslant 0$. For simplicity, we assume isotropic states $\xi=\xi(\pi_{11},\pi_{22},\pi_{33})$ and that the diagonal components are all equal, $\pi_{\cI\cI}=\chi$ for all $\cI$, so that the equation of motion is recast as \cite{Gie13} 
\be\label{gfteomqs4}
2\chi(1-\chi)\frac{\rmd^2\xi(\chi)}{\rmd \chi^2}+(3-4\chi)\frac{\rmd\xi(\chi)}{\rmd \chi}+m\, \xi(\chi)=0\,,
\ee
where $0\leqslant\chi\leqslant 1$ and $m=M^2/12$. To summarize, \emph{homogeneity} is recovered after taking the continuum limit of a special but fully quantum state, while \emph{isotropy} is imposed only to find analytic solutions of the GFT condensate giving rise to the simplest cosmological background. To give $\chi$ an interpretation, we observe that, if the connection remains approximately constant along a dual link with length $\ell_0$, the holonomy thereon is $g\simeq\exp(\ell_0\vec\om\cdot\vec\tau)$, where $\om^i=e^\a A_\a^i$. One has $g=\cos(\ell_0|\vec\om|/2)\,\mathbbm{1}_2+2\vec\tau\cdot(\vec\om/|\vec\om|)\,\sin(\ell_0|\vec\om|/2)$, leading to the identification
\be\label{chisin}
\chi=\sin^2\left(\frac{\ell_0|\vec\om|}{2}\right)=:\sin^2 \left(\frac{\bu c}{2}\right).
\ee
In the second step, we used a notation reminiscent of the LQC cosmological setting, where we encoded the information on the holonomy length and macroscopic fiducial volume into a parameter $\bu=\bu(a)$, whose time evolution is encoded in a scale-factor dependence. Near the identity, $\vec\pi\simeq\vec\om/2$ and $\sqrt{\chi}\simeq \bu c/2$ are proportional to the connection $c$ at low curvature. At the classical level for $\textsc{k}=0$, $c\propto \dot a$, so that the low-curvature classical limit is %($H:=\dot a/a$)
\be\label{chicla}
\chi \propto (a\bu H)^2\ll 1\,,\qquad H:=\dot a/a\,.
\ee

The general solution of Eq.\ \Eq{gfteomqs4}, which we omit \cite{Gie13,CQC}, is always normalizable with respect to the group measure. In the most general case, $\xi(\chi\ll 1)\sim \chi^{-1/2}$ and hence, consistently, the general isotropic vacuum solution is infinitely peaked at small curvature: in the continuum limit, tetrahedra of a classical geometry are nearly flat (spatially constant triad and connection). The exact vacuum solutions of \Eq{gfteomqs4} are well defined also in high-curvature regimes where $\chi\approx 1$ and the flatness condition fails. These regimes are not unphysical, but they do not admit a simple geometric interpretation in the language of continuous smooth manifolds. This situation is strongly remindful of what happens in LQC, where a nonclassical dynamics is effectively encoded in equations on a continuum even if there is no underlying smooth manifold structure. It is in this sense that the universe described by $\xi(\chi)$ is highly quantum, contrary to the WKB wave functions of canonical quantum cosmology which represent conventional semiclassical geometries for all values of their arguments $a$ and $\phi$. 

%%%%%%%%%%%%%%%%%%%%%%%%%%%%%%%%%%%%%%%%%%%%%%%%%%%%%%%%%%%%%%%%%%%%%%%%%%%%%%%%%%%%%%%%%
%%%%%%%%%%%%%%%%%%%%%%%%%%%%%%%%%%%%%%%%%%%%%%%%%%%%%%%%%%%%%%%%%%%%%%%%%%%%%%%%%%%%%%%%%

\section{LQC from GFC} 

We now recover both the classical Friedmann equation and the one of homogeneous and isotropic LQC with a unique parametrization. As in \cite{GOS1}, we take a WKB \emph{Ansatz} of the form
\be\label{xics}
\xi_\textsc{wkb}(\chi,\phi)=\cA(\chi,\phi)\,\rme^{[\rmi S(\chi,\phi)-I(\chi,\phi)]/\lp^2}.
\ee
The functions $\cA$, $S$, and $I$ must be tuned so that the classical-geometry interpretation with $\xi\sim \chi^{-1/2}$ be given by the simultaneous limits $\lp\to 0$ and $\chi\to 0$. Incidentally, the usual problem that WKB states are non-normalizable approximations of unknown quantum-gravity states is solved in GFT, since the general exact solution of \Eq{gfteomqs5} is known.

We look for solutions of \Eq{gfteomqs5} of the form $S=S(\chi,\phi)$ and $I=0$; damping terms can be included in $\cA$. Plugging \Eq{xics} into \Eq{gfteomqs4} (with matter switched on), expanding $m$ as $m=m_4\lp^{-4}+m_2\lp^{-2}+m_1\lp^{-1}+m_0$, and separating order by order in $\lp$, one obtains $O(\lp^{-4})$, $O(\lp^{-2})$, and $O(1)$ equations, of which we report only the first:
\be\label{chichi}
2\chi(1-\chi)(S_{,\chi})^2=-\cE^2(S_{,\phi})^2+m_4\,.
\ee
As in the usual Hamilton--Jacobi formalism, we identify $\p_\chi S\propto p_\chi$ and $\p_\phi S\propto p_\phi$ with the semiclassical momentum conjugate to, respectively, $\chi$ and $\phi$. Classically, in $N=1$ gauge they correspond to
\be\label{pichi}
p_\chi\sim p_{\bu^2\dot a^2}\sim \frac{a}{\bu^2 H}\,,\qquad p_\phi\sim a^3\dot \phi\,.
\ee
Our main results will be that (A) the purely classical limit fixes the behavior of $\bu$ and (B) the limit of LQC effective dynamics is also recovered and confirms (A).

(A) In the low-curvature limit $\chi\ll 1$ but without expanding the left-hand side of \Eq{chichi} as $\chi(1-\chi)\simeq \chi$, using Eq.\ \Eq{chicla} one would have 
%(a\bu H)^2[1-(a\bu H)^2]\propto -\cE^2 p_\phi^2/p_\chi^2 \propto -\cE^2(a\bu)^2(a\bu H)^2\dot\phi^2$, dividing by 2(a\bu)^2(a\bu H)^2 p_\chi^2=2(a^4\bu^4 H^2) a^2/(\bu^4 H^2)=2a^6
 $H^2\propto \cE^2\dot\phi^2-m_4 a^{-6}+(a\bu)^{-2}$, which would be, assuming $\cE^2>0$, the standard Friedmann equation for a massless scalar field and two extra contributions. One is a stiff matter term which can be removed by setting $m_4=0$. The other is a curvature term if $\bu=1$ or a cosmological constant if Eq.~\Eq{gftqs} holds. The first possibility is excluded because the curvature term could only come from the classical connection $c=\g\dot a+1$ and also because, if we want to embed LQC in group field cosmology and identify the GFC function $\bu$ with the LQC function \Eq{bu2}, LQC forbids a constant $\bu$. The other choice is more interesting, but we will see that $\cE^2>0$ does not lead to Lorentzian LQC. Also, Wick rotating the above equation to compensate for a positive $\cE^2$ ($H^2\to-H^2$, $\dot\phi^2\to-\dot\phi^2$) would give a negative cosmological constant.
%after conveniently fixing the proportionality constant
%\be\label{gftfreq}
%H^2= \frac{\k^2}{3}\frac{\dot\phi^2}{2}+\frac{1}{a^2\bu^2}\,.
%\ee
 We therefore turn to another derivation of the classical equation of motion. Taking the extreme regime $\chi\approx 0$, we now make the expansion $\chi(1-\chi)\simeq \chi$ and get
\be\label{gftfreq2}
(a\bu)^{-2}\propto  -\cE^2\dot\phi^2\,.
\ee
If we take $\cE^2<0$, the right-hand side is the scalar field energy density plus a cosmological constant. The left-hand side is $H^2$ only if $\bu\propto 1/\dot a$. For the inverse power law \Eq{bu2} and an expanding universe, this condition is verified if $a\propto t^{1/(1-2n)}$ for $n< 1/2$, or if $a\propto \rme^{Ht}$ when $H$ is constant in the improved quantization scheme $n=1/2$. Although both cases rely on a specific form of the scale factor, the second is more realistic in the presence of a cosmological constant $\Lambda$, which is bound to dominate over matter asymptotically (de Sitter attractor). Remarkably, the choice \Eq{gftqs} is the one of the improved quantization scheme. While in the canonical theory any choice of $\bu$ is formally compatible with the classical limit, here GFC has the correct limit of the Einstein-gravity Friedmann equation in de Sitter approximation only when \Eq{gftqs} holds. There is thus the possibility that part of the ambiguities of the canonical theory be removed in this model of GFT. %At the level of homogeneous and isotropic cosmology, any other $\mu(a)$ in the low-curvature regime would make the embedding of LQC into GFC at least dubious.

(B) We now obtain the Friedmann equation \Eq{Hg2} of LQC for general $\chi$. Observing that $4\chi(1-\chi)=\sin^2(\bu c)$, Eqs.\ \Eq{chichi} and \Eq{Hg2} agree provided $\cE^2<0$ and the Hamilton--Jacobi momentum $p_\chi$ be
\be\label{pichi2}
p_\chi \propto \sqrt{\frac{\a}{\nu}}\frac{a^2}{\bu}\,,
\ee
where $\a$ and $\nu$ are the inverse-volume LQC corrections of the gravity and matter sectors. Equation \Eq{pichi2} has not been derived from first principles, but the characteristic structure of LQC dynamics is indeed reproduced. The classical limit $\a,\nu\to 1$ agrees with Eq.\ \Eq{pichi} only if $\bu\propto 1/\dot a$, consistently with Eq.\ \Eq{gftfreq2}.

The claim that LQC at large can be fully derived from GFT is premature, since our results should be refined in many ways. First, the stability of the condensate should (but can) be checked in perturbation theory, which would be also crucial for the study of cosmological inhomogeneities and their comparison with LQC perturbations. Second, nontrivial interactions $V$ must be turned on to better account for the matter content of the model and obtain a fuller derivation of Eq.~\Eq{pichi2} and the functions $\a$ and $\nu$. Preliminary WKB calculations with the local potential $V= M^2|\vp(g)|^2+(\s/2)|\vp(g)|^4$ show that interactions could generate a nonminimally coupled term. Expanding the coupling $\la:=\s/12=\la_4\lp^{-4}+\la_2\lp^{-2}+\la_0$ and solving the WKB equation $\cA_{,\phi\phi}+(m_0/\cE^2)\cA+(\la_0/\cE^2)\cA^3=0$ for an amplitude $\cA=\cA(\phi)$, the only change in Eq.~\Eq{chichi} is $m_4\to m_4+\la_4\cA^2(\phi)$. The most general analytic solutions are the Jacobi elliptic functions cn, sn, and dn. When $m_0\neq 0$ and assuming $\cE^2<0$, the simplest solutions are $\cA_1(\phi) =\sqrt{-m_0/\la_0}\,\tanh[\sqrt{m_0/(2\cE^2)}\,\phi]=\sqrt{m_0/\la_0}\,\tan[\sqrt{-m_0/(2\cE^2)}\,\phi]$ and $\cA_2(\phi) = \sqrt{-2m_0/\la_0}/\cos(\sqrt{m_0/\cE^2}\,\phi)$, depending on the sign of $m_0$. When $m_0=0$, the only solution is $\cA_3(\phi) = {\rm cn}[\sqrt{\la_0/\cE^2}\,\phi;\,(\sqrt{5}-1)/2]$. The interpretation of these scalar-field profiles with a stiff-matter scaling $a^{-6}$ remains to be assessed and might require to go beyond the WKB approximation.

The present approach will likely have something to say about the cosmological constant problem, too. The peak $\xi\sim \chi^{-1/2}$ in the probability density can be translated into one for $\Lambda$, since in the classical limit with Eq.\ \Eq{gftqs} and for negligible (or nearly constant) matter energy density $\chi\simeq H^2\propto \Lambda$: $\xi(\Lambda)\sim \Lambda^{-1/2}$. This peak is less pronounced than the exponential probabilities found in Wheeler--DeWitt quantum cosmology, but it is perhaps better motivated, as it does not rely on a minisuperspace quantization. 

\section*{Acknowledgments}

I warmly thank M.\ Bojowald, S.\ Gielen, and D.~Oriti for useful discussions. This work is under a Ram\'on y Cajal contract.

\paragraph*{Note added.---} While this work was under completion, we became aware of a forthcoming paper on the same topic \cite{GO}. Where comparable, the results agree. %In particular, the authors of \cite{GO} note that, in GFT, it natural to adopt the lattice-refinement interpretation of LQC \cite{bo609}, where $\bu$ is related to the expectation value of the number of GFT quanta in the fiducial volume.

%%%%%%%%%%%%%%%%%%%%%%%%%%%%%%%%%%%%%%%%%%%%%%%%%%%%%%%%%%%%%%%%%%%%%%%%%%%%%%%%%%%%%%%%%
%%%%%%%%%%%%%%%%%%%%%%%%%%%%%%%%%%%%%%%%%%%%%%%%%%%%%%%%%%%%%%%%%%%%%%%%%%%%%%%%%%%%%%%%%

\end{document}